\documentstyle[seceq,preprint]{ptptex}

\preprintnumber[3cm]{
SMUHEP/99-02}

\title{The Indispensability of Ghost Fields in the Light-Cone \\
 Gauge Quantization of Gauge Fields}

\author{Yuji {\sc Nakawaki} and Gary {\sc McCartor}$^{*}$}

\inst{Division of Physics and Mathematics, Faculty of Engineering,
\\Setsunan University, Osaka 572-8508
\\
$^*$Department of Physics, Southern Methodist University,
\\Dallas TX 75275}

\recdate{\today}

\abst{We continue McCartor and Robertson's recent demonstration of the
indispensability of ghost fields in the light-cone gauge quantization
of gauge fields. It is shown that the ghost fields are indispensable in
deriving well-defined antiderivatives and in regularizing the most
singular component of gauge field propagator. To this end it is
sufficient to confine ourselves to noninteracting abelian fields.
Furthermore to circumvent dealing with constrained systems, we
construct the temporal gauge canonical formulation of the free
electromagnetic field in auxiliary coordinates
$x^{\mu}=(x^-,x^+,x^1,x^2)$ where $x^-=x^0{\rm cos} {\theta}-x^3{\rm
sin}{\theta},\; x^+=x^0{\rm sin}{\theta}+x^3{\rm cos} {\theta}$ and
$x^-$ plays the role of time. In so doing we can quantize the fields
canonically without any constraints, unambiguously introduce "static
ghost fields" as residual gauge degrees of freedom  and construct the
light-cone gauge solution in the light-cone representation by simply
taking the light-cone limit (${\theta}\to\frac{\pi}{4}$). As a by
product we find that, with a suitable choice of vacuum the
Mandelstam-Leibbrandt form of the propagator can be derived in the
${\theta}=0$ case (the temporal gauge formulation in the equal-time
representation).  }

\def\vecx{\mbox {\boldmath $x$}}
\def\vecy{\mbox {\boldmath $y$}}
\def\veck{\mbox {\boldmath $k$}}
\def\vecq{\mbox {\boldmath $q$}}

\begin{document}

\maketitle

\section{Introduction}
Recently the search for nonperturbative solutions of QCD has led to an
extensive explorations of light-front field theory(LFFT), in which
the infinite-momentum limit is incorporated by the change of variables$^{1)}$
 $x^+_l=\frac{x^0+x^3}{\sqrt2},\;x^-_l=\frac{x^0-x^3}{\sqrt2}$ so that one is
able to have vacuum state composed only of particles with nonnegative
longitudinal momentum and also to have relativistic bound-state equations of
Schr\"odinger-type. For a good overview of LFFT see ref.2)

A fundamental problem is to specify the antiderivatives which arise in relating
constrained fields to the true degrees of freedom of the system. Quantization
has traditionally been carried out in parallel with the axial gauge formulation
of QED in the ordinary space-time coordinates. Thus $x^+_l$ and
$A_-^a=\frac{A_0^a-A_3^a}
{\sqrt{2}}=0$ have been chosen respectively to be the evolution parameter
and the gauge fixing condition and the Gauss' law constraint has been solved by
using the operator $({\partial}_-^l)^{-1}$ to express Hamiltonian in
terms of the physical degrees of freedom.$^{3)}$ Unfortunately, the quantity
$(({\partial}_-^l)^{-1})^2$, which is needed for that construction, turns out
to be not well
defined.$^{4)}$ In fact if one defines  $({\partial}_-^l)^{-1}$ by
\begin{equation}
({\partial}_-^l)^{-1}f(x^-_l)=\frac{1}{2}\int_{-{\infty}}^{\infty}dy^-_l
{\epsilon}(x^-_l-y^-_l)f(y^-_l), 
\end{equation}
then $(({\partial}_-^l)^{-1})^2$ is divergent.$^{5)}$ Thus one subtracts
divergent terms by hand to define $({\partial}_-^l)^{-2}$ by
\begin{equation}
({\partial}_-^l)^{-2}f(x^-_l)=\frac{1}{2}\int_{-{\infty}}^{\infty}dy^-_l
|x^-_l-y^-_l|f(y^-_l). 
\end{equation}
Dirac's canonical quantization procedure can not resolve this
difficulty because one has to make use of the same operator $({\partial}_-^l)
^{-1}$to define inverse of the constraints matrix.$^{6)}$

Furthermore, if $(1\cdot1)$ is used to define the inverse derivative,
the spurious singularity at $n \cdot k=0$ of the gauge field propagator
\begin{equation}
 D_{{\mu}{\nu}}^{ab}(k)=\frac{i{\delta}_{ab}}{k^2+i{\epsilon}}
(-g_{{\mu}{\nu}}+\frac{n_{\mu}k_{\nu}+n_{\nu}k_{\mu}}{n \cdot k}) 
\end{equation}
is necessarily defined as the principal value(PV); but evaluating
the singularity  as the PV generates extra contributions so that light-cone
gauge calculations do not agree with those performed in covariant
gauges.$^{7)}$  To overcome the latter difficulty  the
Mandelstam-Leibbrandt(ML)\cite{} prescription has been proposed so as not to
generate extra contributions and give results consistent with Feynman
propagators.$^{8)}$  Bassetto et al derived the ML
prescription in the light-cone gauge using a canonical operator formalism in
the
ordinary space-time coordinates and found that ghost fields associated with
Lagrange
multiplier fields are essential to the derivation.$^{9)}$

Since there exists  an operator solution which has the ML form of the
propagator,
one expects that a consistent operator formulation of LFFT can
also be constructed by introducing the ghost fields as residual gauge degrees
of
freedom.  However, since the residual gauge functions in the
light-cone gauge formulation are ones depending on $x^+_l,x^1$ and $x^2$, one
cannot
expect to calculate the dynamical operators by integrating densities over the
three
dimensional hyperplanes $x^+_l=$constant.
Recently these problems were studied by one of the authors of the present
paper(McCartor) and Robertson in the light-cone formulation
of QED.$^{10)}$ They found that the ghost fields can be introduced in such a
way
that the translational generator $P_+^l$ consists of physical degrees of
freedom
integrated over the hyperplane $x^+_l=$constant and ghost degrees of freedom
integrated over the hyperplane $x^-_l=$constant. They also found that
the ghost fields have to be initialized along a hyperplane $x^-_l=$
constant, while physical fields evolve from the usual hyperplane $x^+_l=$
constant. The same problems were considered by Morara and Soldati$^{11)}$, who
constructed the light-cone temporal gauge formulation, where all fields
evolve from a single initial value surface and the ML form of the propagator is
realized.

In this paper we further investigate how the ghost fields fulfill roles as
regulator fields in the light-cone gauge formulation of gauge theories.
To avoid inessential complications and to circumvent dealing with
constrained systems, we confine ourselves to noninteracting abelian fields ---
the free electromagnetic fields --- and construct the canonical
operator solution of them in the gauge $A_-=A^0{\rm cos}{\theta}+x^3
{\rm sin}{\theta}=0$ and in the auxiliary coordinates $x^{\mu}=
(x^-,x^+,x^1,x^2)$  where
\begin{equation}
x^-=x^0{\rm cos}
{\theta}-x^3{\rm sin}{\theta},\; x^+=x^0{\rm sin}{\theta}+x^3{\rm cos}
{\theta}.
\end{equation}
The same framework was also used by Hornbostel to analyze two-dimensional
models.$^{12)}$ In doing so we can choose $x^-$ as the evolution parameter in
the interval $0{\le}{\theta}<\frac{\pi}{4}$ and construct the temporal gauge
formulation, where canonical quantization conditions are to be imposed without
any constraints. Furthermore we can take $x^+$ to be the evolution parameter
in the interval $\frac{\pi}{4}<{\theta}{\le}\frac{\pi}{2}$ and thus construct
the axial gauge formulation. Consequently we can expect that taking the
light-cone limit ${\theta}\to\frac{\pi}{4}$ will enable us to find light-cone
gauge solutions and to compare the temporal light-front limit
(${\theta}\to\frac{\pi}{4}-0$) with the axial light-front
limit(${\theta}\to\frac{\pi}{4}+0$).

In sect.2 the temporal gauge canonical operator solution is constructed
in the auxiliary coordinates and it is shown that static ghost fields are
present as residual gauge degrees of freedom. We also encounter the problem
that canonical commutation relations can not distinguish the PV and ML
prescriptions$^{13)}$ but that the ML
prescription can always be obtained if we employ an appropriate representation
of the ghost fields. As a consequence we find that the ML prescription
can always be implemented even in the ${\theta}=0$ case, namely in the temporal
gauge formulation in the ordinary space-time coordinates. We also point
out that the axial gauge formulation is not straightforward to construct, at
least by these procedures.

In sect.3 we show that the free electromagnetic fields given as the
temporal light-cone limit are identical with the ones given by McCartor and
Robertson and by Morara and Soldati and equivalent to those given by Bassetto
et al in the ordinary space-time coordinates. We also show in detail that in
the
light-cone gauge formulation the ghost fields are indispensable
to well-defined antiderivatives and that linear divergences are
eliminated in what is otherwise the most singular component of $x_l^+$- ordered
propagator.

Sect.4 is devoted to concluding remarks.

\section{Temporal Gauge Formulation in the Auxiliary Coordinates}

We begin by fixing the metric of the auxiliary coordinates $x^{\mu}=
(x^-,x^+,x^1,x^2)$ where $x^-$ and $x^+$ are defined by
\begin{equation}
x^-=x^0{\rm cos}{\theta}-x^3{\rm sin}{\theta},\quad
x^+=x^0{\rm sin}{\theta}+x^3{\rm cos}{\theta}.
\end{equation}
Inverting these, we find that $x^0$ and $x^3$ are given by
\begin{equation}
x^0=x^-{\rm cos}{\theta}+x^+{\rm sin}{\theta},\quad
x^3=-x^-{\rm sin}{\theta}+x^+{\rm cos}{\theta}
\end{equation}
so that  $(x^0)^2-(x^3)^2$ is expressed in terms of $x^-$ and $x^+$ as
\begin{equation}
(x^0)^2-(x^3)^2=x^-(x^-{\rm cos}2{\theta}+x^+{\rm sin}2{\theta})
+x^+(x^-{\rm sin}2{\theta}-x^+{\rm cos}2{\theta}).
\end{equation}
Rewriting this in the form $x^-x_-+x^+x_+$ requires that
$x_-$ and $x_+$ are defined respectively by
\begin{equation}
x_-=x^-{\rm cos}2{\theta}+x^+{\rm sin}2{\theta},\quad
x_+=x^-{\rm sin}2{\theta}-x^+{\rm cos}2{\theta}. 
\end{equation}
It follows from this that
\begin{equation}
g_{--}={\rm cos}2{\theta},\;g_{-+}=g_{+-}={\rm sin}2{\theta},
\;g_{++}=-{\rm cos}2{\theta}. 
\end{equation}
Upper components of metric tensor are obtained by inverting $(2\cdot4)$
 \begin{equation}
g^{--}={\rm cos}2{\theta},\;g^{-+}=g^{+-}={\rm sin}2{\theta},
\;g^{++}=-{\rm cos}2{\theta}.
\end{equation}
Substituting $(2\cdot1)$ into $(2\cdot4)$ enables us to
express $x_-$ and $x_+$ in terms of $x^0$ and $x^3$ as follows
\begin{equation}
x_-=x^0{\rm cos}{\theta}+x^3{\rm sin}{\theta},\quad
x_+=x^0{\rm sin}{\theta}-x^3{\rm cos}{\theta}.
\end{equation}

Now we notice from $(2\cdot5)$ and $(2\cdot6)$ that if we keep ${\theta}$
in the interval $0{\le}{\theta}<\frac{\pi}{4}$, we can employ $x^-$ as
an evolution parameter, whereas in the interval $\frac{\pi}{4}<{\theta}{\le}
\frac{\pi}{2}$ we can employ $x^+$ as an evolution parameter. Thus we
expect that fixing ${\theta}$ to be $\frac{\pi}{4}$ will allow us to have
either
$x^-$ or $x^+$ as evolution parameters.

\subsection{Temporal Gauge Quantization}
To construct the temporal gauge formulation in the auxiliary coordinates, we
choose  $A_-=A_0{\rm cos}{\theta}-A_3{\rm sin}{\theta}=0$ as the gauge fixing
condition. Accordingly  we consider the Lagrangian
\begin{equation}
L=-\frac{1}{4}F_{{\mu}{\nu}}F^{{\mu}{\nu}}-B(n \cdot A) 
\end{equation}
where $F_{{\mu}{\nu}}={\partial}_{\mu}A_{\nu}- {\partial}_{\nu}A_{\mu}$
 with ${\partial}_{\mu}=(\frac{{\partial}}{{\partial}x^-},
\;\frac{{\partial}}{{\partial}x^+},\;\frac{{\partial}}{{\partial}x^1},\;
\frac{{\partial}}{{\partial}x^2}), \;n^{\mu}=(n^-,n^+,n^1,n^2)=(1,0,0,0)$
and $B$ is a Lagrange multiplier field. From $(2\cdot8)$ we can derive the
field
equations
\begin{equation}
{\partial}_{\mu}F^{{\mu}{\nu}}={\Box}A^{\nu}-{\partial}^{\nu}
({\partial}^{\mu}A_{\mu})=n^{\nu}B  
\end{equation}
and the gauge fixing condition
\begin{equation}
A_-=0. 
\end{equation}
The field equation of $B$,
\begin{equation}
{\partial}_-B=0  
\end{equation}
is obtained by multiplying $(2\cdot9)$ by ${\partial}_{\nu}$.
Lowering the indices of $(2\cdot9)$ and observing that $n_{\mu}=({\rm cos}
2{\theta},{\rm sin}2{\theta},0,0)$ and $A_-=0$ enable us to derive the field
equation of $A_-$ as
\begin{equation}
{\partial}_-({\partial}^{\mu}A_{\mu})+{\rm cos}2{\theta}B=0.  
\end{equation}
Then, differentiating this by ${\partial}_-$ gives rise to the following
equation for ${\partial}^{\mu}A_{\mu}$
\begin{equation}
{\partial}_-^2({\partial}^{\mu}A_{\mu})=0. 
\end{equation}
Consequently, upon multiplying $(2\cdot9)$ by ${\partial}_-^2$ we obtain the
following equation for $A_{\mu}$
\begin{equation}
{\Box}({\partial}_-^2A_{\mu})=0. 
\end{equation}

>From $(2\cdot8)$ we can also obtain the canonical conjugate momenta
\begin{equation}
{\pi}^-=0, \quad {\pi}^+=-F^{-+},\quad {\pi}^i=-F^{-i}, \quad {\pi}_B=0.
\end{equation}
It should be noticed that we have three pairs of canonical variables, in
contrast with one pair in the light-cone temporal gauge formulation.
The fields $B,\;{\partial}_ -A_+$ and ${\partial}_- A_i\;(i=1,2)$ are
expressed in terms of canonical variables as follows
\begin{equation}
B={\partial}_+{\pi}^++\sum_{i=1}^2{\partial}_i{\pi}^i, \quad
{\partial}_- A_+={\pi}^+, \quad
{\partial}_- A_i=\frac{{\pi}^i-{\rm sin}2{\theta}\;F_{+i}}{{\rm cos}2{\theta}}.
\end{equation}
Thus we can impose the following equal-time canonical quantization conditions
\begin{equation}
[A_r(x),\;A_s(y)]=0, \quad [A_r(x),\;{\pi}^s(y)]=i{\delta}_{rs}{\delta}^{(3)}
(\vecx^+-\vecy^+),\quad [{\pi}^r(x),\;{\pi}^s(y)]=0 
\end{equation}
where $r,s=+,1,2,\;{\delta}^{(3)}(\vecx^+-\vecy^+){\equiv}{\delta}(x^1-y^1)
{\delta}(x^2-y^2){\delta}(x^+-y^+)$ and in all cases $x^-=y^-$.

To obtain 4-dimensional commutation relations of $A_{\mu}$ we express
$A_{\mu}$  in an integral form.$^{14)}$ This is done most easily by
making use of the commutator function of a free massless field
 \begin{equation}
D(x)=\frac{-i}{2(2{\pi})^3}\int\frac{d^3k_+}{k^-}({\rm e}^{-ik \cdot x}-
{\rm e}^{ik \cdot x}) 
\end{equation}
and a commutator function $E(x)$ satisfying the equation
\begin{equation}
{\Box}E(x)=-x^-{\delta}^{(3)}(\vecx^+){\equiv}D_s(x), \quad
{\partial}_-^2E(x)=D(x) 
\end{equation}
and the initial conditions
\begin{equation}
E(x)|_{x^-=0}=0, \, {\partial}_-E(x)|_{x^-=0}=0,
\; {\partial}_-^2E(x)|_{x^-=0}=0, \;
{\partial}^-{\partial}_-^2E(x)|_{x^-=0}
=-{\delta}^{(3)}(\vecx^+). 
\end{equation}
Note that $D(x)$ satisfies a free massless D'Alember's equation, which imposes
the following on-shell condition on four momentum $k_{\mu}$
\begin{equation}
0=k^2=n^2k_-^2+2{\rm sin}2{\theta}k_+k_--k_1^2-k_2^2-n^2k_+^2, \quad(n^2=
{\rm cos}2{\theta}). 
\end{equation}
We solve this in such a way that $k_-$ is expressed in terms of $k_1,\;k_2,\;
k_+$ as follows
\begin{equation}
k_-=\frac{k_{\bot}^2+n^2k_+^2}{k^-+{\rm sin}2{\theta}k_+} 
\end{equation}
where
\begin{equation}
k_{\bot}=\sqrt{k_1^2+k_2^2}, \quad k^-=\sqrt{k_+^2+n^2k_{\bot}^2}. 
\end{equation}
Also note that $d^3k_+$ denotes $dk_1dk_2dk_+$ and the integration region of
$k_+$ is $(-{\infty},\;{\infty})$.

We find that the function $E(x)$ is given by
\begin{equation}
E(x)=\frac{1}{{\partial}_-^2}D(x)-\frac{1}{{\partial}_{\bot}^2+
n^2{\partial}_+^2}
D_s(x)-\frac{{\rm sin}2{\theta}{\partial}_++{\partial}^-}
{({\partial}_{\bot}^2+n^2{\partial}_+^2)^2}{\partial}_-D_s(x) 
\end{equation}
where
\begin{equation}
\frac{1}{{\partial}_-}{\equiv}\frac{{\rm
sin}2{\theta}{\partial}_++{\partial}^-}
{{\partial}_{\bot}^2+n^2{\partial}_+^2}, \quad
{\partial}_{\bot}^2{\equiv}{\partial}_1^2+{\partial}_2^2. 
\end{equation}
Now we can express $A_{\mu}$ in the following integral form
\begin{equation}
A_{\mu}(x)=\!\int
d^3z^+[{\partial}_z^-D(x-z)A_{\mu}(z)-D(x-z){\partial}^-A_{\mu}(z)
+{\partial}_-^zE(x-z){\Box}A_{\mu}(z)-E(x-z){\Box}{\partial}_-A_{\mu}(z) ]
\end{equation}
where $d^3z^+$ denotes $dz^1dz^2dz^+$. It can easily be shown that the
integral form satisfies the field equation ${\Box}({\partial}_-^2A_{\mu})=0$
and is independent of $z^-$ so that it satisfies the initial conditions at
$z^-=x^-$.
Furthermore, we can utilize the latter property to calculate its 4-dimensional
commutation relations using only equal-time canonical commutation
relations. It turns out that
\begin{equation}
[A_{\mu}(x),\;A_{\nu}(y)]=i\{ -g_{{\mu}{\nu}}D(x-y)+(n_{\mu}{\partial}_{\nu}
+ n_{\nu}{\partial}_{\mu}){\partial}_-E(x-y)-n^2{\partial}_{\mu}
{\partial}_{\nu}E(x-y) \}. 
\end{equation}
>From $(2\cdot12)$, $B$ is expressed as
\begin{equation}
B=-\frac{{\partial}_-({\partial}^{\mu}A_{\mu})}{{\rm cos}2{\theta}} 
\end{equation}
so the commutation relations of $B$ are obtained from $(2\cdot27)$ as follows
\begin{equation}
[B(x),\;A_{\nu}(y)]=-i{\partial}_{\nu}{\delta}^{(3)}(\vecx^+-\vecy^+),
\quad [B(x),\;B(y)]=0. 
\end{equation}

\subsection{Constituent fields in the Temporal Gauge Formulation}
To obtain constituent fields in the temporal gauge formulation, we solve
the field equations $(2\cdot9)$. We multiply $(2\cdot9)$ with
${\nu}=i\;(i=1,2)$ by ${\partial}_i$ and sum over $i=1,2$. Consequently we
obtain
\begin{equation}
{\partial}^{\mu}A_{\mu}=\frac{\Box}{{\partial}_{\bot}^2}\sum_{i=1}^2
{\partial}_iA_i 
\end{equation}
so that  $(2\cdot9)$ is rewritten as
\begin{equation}
{\Box}(A_{\mu}-\frac{{\partial}_{\mu}}{{\partial}_{\bot}^2}\sum_{i=1}^2
{\partial}_iA_i)=n_{\mu}B. 
\end{equation}
Because $B$ satisfies ${\partial}_-B=0$, any solution of $(2\cdot31)$ is
described as
\begin{equation}
A_{\mu}-\frac{{\partial}_{\mu}}{{\partial}_{\bot}^2}\sum_{i=1}^2
{\partial}_iA_i=a_{\mu}-\frac{n_{\mu}}{{\partial}_{\bot}^2+n^2{\partial}_+^2}B
\end{equation}
where $a_{\mu}$ is a homogeneous solution. As a condition that $A_{\mu}$ in
$(2\cdot32)$ satisfies $(2\cdot30)$, we obtain
\begin{equation}
{\partial}^{\mu}a_{\mu}=0.
\end{equation}
Furthermore multiplying $(2\cdot32)$ with ${\mu}=i\;(i=1,2)$ by ${\partial}_i$
and summing over $i=1,2$ yields another consistency condition
\begin{equation}
\sum_{i=1}^2{\partial}_ia_i=0.
\end{equation}
To obtain $\sum_{i=1}^2{\partial}_iA_i$, we impose the gauge fixing condition
on $(2\cdot32)$ with ${\mu}=-$ and integrate it with respect to $x^-$.
It should be noted here that we can introduce a static field, which we denote
$C$,
as an integration constant.  For later convenience we introduce it in the
following way
\begin{equation}
-\frac{1}{{\partial}_{\bot}^2}\sum_{i=1}^2
{\partial}_iA_i=\frac{1}{{\partial}_-}a_--\frac{n_-x^-}{{\partial}_{\bot}^2+n^2
{\partial}_+^2}B +\frac{1}{{\partial}_{\bot}^2+n^2{\partial}_+^2}
(C-\frac{n^2{\rm sin}2{\theta}{\partial}_+}{{\partial}_{\bot}^2+n^2
{\partial}_+^2}B). 
\end{equation}
On substituting $(2\cdot35)$ into $(2\cdot32)$ we have
\begin{equation}
A_{\mu}=a_{\mu}-\frac{{\partial}_{\mu}}{{\partial}_-}a_-+{\Gamma}_{\mu}
\end{equation}
where
\begin{equation}
{\Gamma}_{\mu}=-\frac{n_{\mu}}{{\partial}_{\bot}^2+n^2{\partial}_+^2}B
-\frac{{\partial}_{\mu}}{{\partial}_{\bot}^2+n^2{\partial}_+^2}
(C-n^2x^-B-\frac{n^2{\rm sin}2{\theta}{\partial}_+}{{\partial}_{\bot}^2+n^2
{\partial}_+^2}B).
\end{equation}

Now that we have obtained the constituent fields $a_{\mu},\;B$ and $\;C,$ we
proceed
to derive the commutation relations they satisfy. It is straightforward to
accomplish if we express them in terms of the canonical variables as follows
\begin{equation}
a_r=A_r-\frac{{\partial}_r}{{\partial}_{\bot}^2}(\sum_{i=1}^2{\partial}_iA_i)
+\frac{n_r}{{\partial}_{\bot}^2+n^2{\partial}_+^2}B, \quad (r=+,1,2)
\end{equation} 
\begin{equation}
B={\partial}_+{\pi}^++\sum_{i=1}^2{\partial}_i{\pi}^i,
\end{equation} 
\begin{equation}
{\partial}^-a_i={\pi}^i-\frac{{\partial}_i}{{\partial}_{\bot}^2}
(\sum_{j=1}^2{\partial}_j{\pi}^j), \quad (i=1,2)
\end{equation} 
\begin{equation}
{\partial}^-a_+=n_-{\pi}^+-\frac{{\partial}_+}{{\partial}_{\bot}^2}
(\sum_{i=1}^2{\partial}_i{\pi}^i)+\frac{n_+^2}{{\partial}_{\bot}^2+n^2
{\partial}_+^2}{\partial}_+B,
\end{equation} 
\begin{equation}
C=n_+{\pi}^+-n_-{\partial}_+A_+-\sum_{i=1}^2{\partial}_iA_i+n^2x^-B-
\frac{n^2n_+}{{\partial}_{\bot}^2+n^2{\partial}_+}{\partial}_+B. \quad
\end{equation}
We find that $a_r,\;{\partial}^-a_r,B,C$ are fundamental operators,
satisfying the following commutation relations
\begin{equation}
[a_r(x),\;a_s(y)]_{x^-=y^-}=
[{\partial}^-a_r(x),\;{\partial}^-a_s(y)]_{x^-=y^-}=0, \quad (r,s =+, 1,2)
\end{equation}
\begin{equation}
[a_+(x),\;{\partial}^- a_+(y)  ]_{x^- = y^-}
=i\frac{{\partial}_+^2+n^2{\partial}_{\bot}^2}{{\partial}_{\bot}^2}
{\delta}^{(3)}(\vecx^+-\vecy^+),  
\end{equation}
\begin{equation}
[a_i(x),\;{\partial}^- a_j(y)  ]_{x^- = y^-}=
i({\delta}_{ij}-\frac{{\partial}_i{\partial}_j}{{\partial}_{\bot}^2})
{\delta}^{(3)}(\vecx^+-\vecy^+), \quad (i,j = 1,2)
\end{equation}
\begin{equation}
[B(x), B(y)]=[C(x), C(y)]=0,
\end{equation} 
\begin{equation}
[B(x), C(y)]=-[C(x), B(y)]=
-i({\partial}_{\bot}^2+n^2{\partial}_+^2){\delta}^{(3)}(\vecx^+-\vecy^+).
\end{equation}
Any other commutators among $a_r,\;{\partial}^-a_r,\;B$ or $\;C$ are zero.

\subsection{Expression of the Constituent fields and
Implementation of the ML prescription}
Now we can express the constituent fields in terms of creation and
annihilaton operators. First of all we notice that the free massless field
$a_+$ satisfies the commutation relation $(2\cdot44)$ with the operator
${\partial}_+^2+n^2{\partial}_{\bot}^2$, which in effect multiplies the
integrand
of  the Fourier expanded free massless fields by $(k^-)^2$. Thus we express
$a_+$
in the form
\begin{equation}
a_+(x) =\frac{-1}{\sqrt{2(2{\pi})^3}}\int \frac{d^3k_+}{\sqrt{k^-}}
\frac{k^-}{k_{\bot}}\{
a_1(\veck_+)e^{-ik\cdot x}+
a_1^{\dagger}(\veck_+)e^{ik\cdot x}\}. 
\end{equation}
Here the factor $-1$ is included for later convenience and $\veck_+$
denotes the three-component vector$(k_+,\;k_1,\;k_2)$. The expression of
$a_-$ is obtained by solving the constraint ${\partial}^+a_++{\partial}^-a_-
=0$ and hence we have
\begin{equation}
a_-(x) =\frac{1}{\sqrt{2(2{\pi})^3}}\int \frac{d^3k_+}{\sqrt{k^-}}
\frac{k^+}{k_{\bot}}\{
a_1(\veck_+)e^{-ik\cdot x}+
a_1^{\dagger}(\veck_+)e^{ik\cdot x}\}. 
\end{equation}
Similarly, because $a_i$ satisfies the constraint
$\sum_{i=1}^2{\partial}_ia_i=0$ and the commutation relation $(2\cdot45)$,
we have
\begin{equation}
a_i(x) =\frac{1}{\sqrt{2(2{\pi})^3}}\int \frac{d^3k_+}{\sqrt{k^-}}
{\epsilon}_i^{({2})}(k) \{
a_2(\veck_+)e^{-ik\cdot x}+ a_2^{\dagger}(\veck_+)e^{ik\cdot x}\}
\end{equation}
where ${\epsilon}_i^{({2})}(k)$ is a physical polarization vector given by
\begin{equation}
{\epsilon}_{\mu}^{(2)}(k)=(0,\;0,\;-\frac{k_2}{k_{\bot}},\;
\frac{k_1}{k_{\bot}}). 
\end{equation}
The operators $a_{{\lambda}}(\veck_+)$ and $a_{\lambda}^{\dagger}
(\veck_+)\;({\lambda}=1,2)$ are normalized so as to satisfy the usual
commutation
relations
\begin{equation}
[a_{\lambda}(\veck_+),\;a_{{\lambda}^{\prime}}(\vecq_+)]=0, \quad
[a_{\lambda}(\veck_+),\;a_{{\lambda}^{\prime}}^{\dagger}(\vecq_+)]=
{\delta}_{{\lambda}{\lambda}^{\prime}}{\delta}^{(3)}(\veck_+-\vecq_+).
\end{equation}
They are nothing but annihilation and creation
operators of physical photons, as we see from the fact that,
with the help of $(2\cdot51)$ and another physical polarization vector
\begin{equation}
{\epsilon}_{\mu}^{(1)}(k)=(0,\;-\frac{k_{\bot}}{k_-},\;-\frac{k^+k_1}
{k_-k_{\bot}},\;-\frac{k^+k_2}{k_-k_{\bot}}), 
\end{equation}
we can express
$a_{\mu}-\frac{{\partial}_{\mu}}
{{\partial}_-}a_-$ in the following compact form
\begin{equation}
a_{\mu}(x)-\frac{{\partial}_{\mu}}{{\partial}_-}a_-(x)
=\frac{1}{\sqrt{2(2{\pi})^3}}\int \frac{d^3k_+}{\sqrt{k^-}}
\sum_{{\lambda}=1}^2{\epsilon}_{\mu}^{({\lambda})}(k) \{
a_{\lambda}(\veck_+)e^{-ik\cdot x}+h.c. \}.
\end{equation}
Note that the polarization vectors satisfy
\begin{equation}
k^{\mu}{\epsilon}_{\mu}^{({\lambda})}(k)=0, \quad
n^{\mu}{\epsilon}_{\mu}^{({\lambda})}(k)=0, \quad ({\lambda}=1,2) 
\end{equation}
\begin{equation}
\sum_{{\lambda}=1}^2{\epsilon}_{\mu}^{({\lambda})}(k)
{\epsilon}_{\nu}^{({\lambda})}(k)=-g_{{\mu}{\nu}}+\frac{n_{\mu}k_{\nu}+
n_{\nu}k_{\mu}}{k_-}-n^2\frac{k_{\mu}k_{\nu}}{k_-^2}. 
\end{equation}

Let us next determine expressions for $B$ and $C$. We expand $B$ in terms of
conjugate zero-norm creation and annihilation operators
$b(\veck_+),\;b^{\dagger}
(\veck_+)$. In so doing we can realize Gauss's
law in physical space specified below. Similarly, we expand
$C$ in terms of conjugate zero-norm creation and annihilation operators
$ c(\veck_+),\;c^{\dagger}(\veck_+)$.
It seems at first sight that there arises no problem in expressing $B$ and
$C$ in terms of the zero-norm operators, because one can employ the following
naive expressions for the static fields
\begin{equation}
B(x) =\frac{1}{\sqrt{2(2{\pi})^3}}\int \frac{d^3k_+}{\sqrt{|k_+|}}
(k_{\bot}^2+n^2k_+^2) \{
b(\veck_+)e^{-ik\cdot x}+
b^{\dagger}(\veck_+)e^{ik\cdot x}\}_{x^-=0}, 
\end{equation}
\begin{equation}
C(x) =\frac{i}{\sqrt{2(2{\pi})^3}}\int d^3k_+{\sqrt{|k_+|}}
 \{ c(\veck_+)e^{-ik\cdot x}-
c^{\dagger}(\veck_+)e^{ik\cdot x}\}_{x^-=0} 
\end{equation}
where
\begin{equation}
[b(\veck_+),\;c^{\dagger}(\vecq_+)]=[c(\veck_+),\;b^{\dagger}(\vecq_+)]=
-{\delta}^{(3)}(\veck_+-\vecq_+), 
\end{equation}
any other commutators being zero. However, we encounter the problem pointed out
by Haller$^{13)}$ , namely the problem that the canonical commutation
relations can not distinguish the PV and ML prescriptions. In fact if we take
the $k_+$-integration region to be the whole interval $(-{\infty},{\infty})$,
then we are obliged to have the PV form of propagator. Judging from the fact
that the ML form of propagator has to be employed in the light-cone limit and
from indications that the PV form of propagator does not lead to the correct
behavior of the Wilson loop in perturbative calculations,$^{15)}$ but
the ML form of propagator does,$^{16)}$ we conclude that the correct form of
temporal gauge free theory from which to start perturbative calculations should
perhaps have the ML prescription in the temporal gauge free propagator.
So far extensions of the ML prescription outside the light-cone gauge
formulation have been done by limiting the $k_+$-integration region by hand to
be $(0,{\infty}).^{17)}$

We solve this problem by
introducing an inequivalent Fock space. We can rewrite
$B$ and $C$ in such a way that $k_+$-integrations are carried over the
interval$(0,\;{\infty})$ as in the following
\begin{equation}
B(x) =\frac{1}{\sqrt{(2{\pi})^3}}\int \frac{d^3k_+}
{\sqrt{k_+}}{\theta}(k_+)(k_{\bot}^2+n^2k_+^2)
\{ B(\veck_+)e^{-ik\cdot x}+
B^{\dagger}(\veck_+)e^{ik\cdot x}\}_{x^-=0},
\end{equation}
\begin{equation}
C(x) =\frac{i}{\sqrt{(2{\pi})^3}}\int d^3k_+{\theta}(k_+){\sqrt{k_+}}
 \{ C(\veck_+)e^{-ik\cdot x}- C^{\dagger}(\veck_+)e^{ik\cdot x}\}_{x^-=0}
\end{equation}
where
\begin{equation}
B(\veck_+)=\frac{b(\veck_+)+b^{\dagger}(-\veck_+)}{\sqrt{2}},\;
B^{\dagger}(\veck_+)=\frac{b^{\dagger}(\veck_+)+b(-\veck_+)}{\sqrt{2}}, 
\end{equation}
\begin{equation}
C(\veck_+)=\frac{c(\veck_+)-c^{\dagger}(-\veck_+)}{\sqrt{2}},\;
C^{\dagger}(\veck_+)=\frac{c^{\dagger}(\veck_+)-c(-\veck_+)}{\sqrt{2}}. 
\end{equation}
Now we see that $B(\veck_+)$ and $C(\veck_+)$ are nothing but canonical
transformations, which are generated as follows
\begin{equation}
B(\veck_+)=e^{\Lambda}b(\veck_+)e^{-\Lambda},\;
B^{\dagger}(\veck_+)=e^{\Lambda}b^{\dagger}(\veck_+)e^{-\Lambda}, 
\end{equation}
\begin{equation}
C(\veck_+)=e^{\Lambda}c(\veck_+)e^{-\Lambda},\;
C^{\dagger}(\veck_+)=e^{\Lambda}c^{\dagger}(\veck_+)e^{-\Lambda}
\end{equation}
where
\begin{equation}
{\Lambda}=\frac{{\pi}}{4}\int d^3k_+{\theta}(k_+)\{ b(\veck_+)c(-\veck_+)
-b(-\veck_+)c(\veck_+)- b^{\dagger}(\veck_+)c^{\dagger}(-\veck_+)+
b^{\dagger}(-\veck_+)c^{\dagger}(\veck_+) \}. 
\end{equation}
Thus, if we define the physical vacuum state $|{\rm{\Omega}}>$  by
\begin{equation}
|{\rm{\Omega}}>=e^{\Lambda}|0> 
\end{equation}
where $|0>$ is the bare vacuum state satisfying
 \begin{equation}
b(\veck_+)|0>=c(\veck_+)|0>=0,  
\end{equation}
then it is easy to show that
\begin{equation}
B(\veck_+)|{\rm{\Omega}}>=C(\veck_+)|{\rm{\Omega}}>=0.  
\end{equation}
Thus physical space ${\rm V_P}$ is defined by
\begin {equation}
{\rm V_P}=\{\; |{\rm phys}>\; | \;B(\veck_+)|{\rm phys}>=0\; \}. 
\end{equation}
In this way we can always obtain the ML prescription in the physical space.
To be complete we note that the $x^-$-ordered propagator
\begin {equation}
D^-_{{\mu}{\nu}}(x-y)=<{\rm{\Omega}}|\{{\theta}(x^--y^-)A_{\mu}(x)A_{\nu}(y)
+{\theta}(y^--x^-)A_{\nu}(y)A_{\mu}(x) \}|{\rm{\Omega}}> 
\end{equation}
results in the ML form of propagator as follows
\begin {equation}
D^-_{{\mu}{\nu}}(x-y)=\frac{1}{(2{\pi})^4}\int d^4kD^-_{{\mu}{\nu}}(k)
e^{-ik \cdot (x-y)}  
\end{equation}
where
\begin{equation}
D^-_{{\mu}{\nu}}(k)=\frac{i}{k^2+i{\epsilon}}\{ -g_{{\mu}{\nu}}
+\frac{n_{\mu}k_{\nu}+n_{\nu}k_{\mu}}{k_-+i{\epsilon}{\rm sgn}(k_+)}
-\frac{n^2k_{\mu}k_{\nu}}{(k_-+i{\epsilon}{\rm sgn}(k_+))^2 } \}. 
\end{equation}
It is known that for interacting theories the ML form of propagator has
to be employed in the  light-cone limit(${\theta}\to\frac{\pi}{4}$).
Therefore it is interesting to investigate whether it is also true of the
${\theta}=0$ case, that is to say, the temporal gauge formulation in the
ordinary space-time coordinates.

Translation generators
$P_{\mu}=\int d^3x^+T_{\mu}^-(x)$ are given in terms of the creation
and annihilation operators of the constituent fields as follows
\begin{equation}
P_-=\int d^3k_+ \{ \sum_{{\lambda}=1}^2
k_-a_{\lambda}^{\dagger}(\veck_+)a_{\lambda}(\veck_+)
+n^2\frac{k_{\bot}^2+n^2k_+^2}{k_+}{\theta}(k_+)
B^{\dagger}(\veck_+)B(\veck_+) \}, 
\end{equation}
\begin{equation}
P_r=\int d^3k_+k_r \{ \sum_{{\lambda}=1}^2
a_{\lambda}^{\dagger}(\veck_+)a_{\lambda}(\veck_+)
-{\theta}(k_+)\left( B^{\dagger}(\veck_+)C(\veck_+)+C^{\dagger}(\veck_+)
B(\veck_+)\right) \},
\end{equation}
where $r=+,1,2$.

We close this section by making two remarks.  First, that the Heisenberg
equation of $C$ is not satisfied, as is seen from
\begin{equation}
[P_-,C(x)]=in^2B(x)
\end{equation}
but that this is necessary to assure the Heisenberg equations of $A_{\mu}$
\begin{equation}
[P_-,A_{\mu}(x)]=-i{\partial}_-A_{\mu}(x).
\end{equation}
Second, that we have not attempted constructing the axial gauge formulation
in the auxiliary coordinates. This is because we have not succeeded in finding
any solutions other than $(2\cdot36)$. It seems to us that $(2\cdot36)$
is not an appropriate solution in the axial gauge formulation because
$x^-$ is included explicitly and because the inverse Laplace operator
$\frac{1}{{\partial}_{\bot}^2+n^2{\partial}_+^2}$ becomes singular.

\section{Light-Cone Gauge Quantizations in the Light-Front Coordinates}
In this section we confine ourselves to the ${\theta}=\frac{\pi}{4}$ case,
namely, to the formulation in the light-front coordinates
$x^{\mu}=(x^-,\;x^+,\;x^1,\;x^2)=(\frac{x^0-x^3}{\sqrt{2}},\;
\frac{x^0+x^3}{\sqrt{2}},\;x^1,\;x^2)$ \footnote{To avoid inessential
complications,
in this section we omit the suffix {\it l} denoting quantities in the
light-front
coordinates and use the same notation as that used in sect.2 .}
with metric tensors defined by
\begin{eqnarray}
g_{11}=g_{22}=g^{11}=g^{22}=-1,  \nonumber \\
g_{-+}=g_{+-}=g^{-+}=g^{+-}=1, \\ 
{\rm all}\;{\rm other}\;{\rm components}=0. \nonumber
\end{eqnarray}

First of all we recall that the Lagrangian $(2\cdot8)$
 becomes singular in the light-front coordinates, where $n_{\mu}=
(0,\;1,\;0,\;0)$. In fact in the
temporal gauge formulation with $x^-$ being the evolution parameter, it
happens that the canonical momenta ${\pi}^i$ conjugate to $A_i$ becomes
noninvertible, as we seen from
\begin{equation}
{\pi}^-=0,\;{\pi}^+={\partial}_-A_+,\;{\pi}^i={\partial}_+A_i-{\partial}_iA_+,
\;(i=1,2),\; {\pi}_B=0. 
\end{equation}
What is worse, in the axial gauge formulation with $x^+$ being the evolution
parameter, all momenta become noninvertible as in the following
\begin{equation}
{\pi}^-=-{\partial}_-A_+, \;{\pi}^+=0,\;
{\pi}^i={\partial}_-A_i,\;(i=1,2),\; {\pi}_B=0. 
\end{equation}

As to the temporal case, we can circumvent dealing with the constrained
system because we can obtain canonically quantized free electromagnetic fields
by simply taking the limit ${\theta}\to\frac{\pi}{4}-0$ of $(2\cdot36)$.
As to the axial case, we can not follow the same approach because we have
not succeeded in finding an appropriate axial gauge solution in the auxiliary
coordinates. Nevertheless, we  expect to get the axial light-front
limit(${\theta}\to\frac{\pi}{4}+0$) from the temporal light-front
limit(${\theta}\to\frac{\pi}{4}-0$).  As a matter of fact, known light-cone
gauge solutions are equivalent to each other, which is seen below.

In the temporal light-front limit $n^2={\rm cos}2{\theta}$ becomes zero and
the mass-shell condition of the free massless fields is changed into
$2k_-k_+-k_{\bot}^2=0$ so that the range of $k_+$-integration is reduced
to $(0,{\infty})$ in the Fourier expansions of the free massless fields.
Consequently we obtain the following electromagnetic fields described in
terms of the constituent fields
\begin{equation}
A_{\mu}=a_{\mu}-\frac{{\partial}_{\mu}}{{\partial}_-}a_-
 -\frac{n_{\mu}}{{\partial}_{\bot}^2}B-
\frac{{\partial}_{\mu}}{{\partial}_{\bot}^2}C 
\end{equation}
where
\begin{equation}
a_+(x) =\frac{-1}{\sqrt{2(2{\pi})^3}}\int \frac{d^3k_+}{\sqrt{k_+}}
\frac{k_+}{k_{\bot}}{\theta}(k_+)\{
a_1(\veck_+)e^{-ik\cdot x}+
a_1^{\dagger}(\veck_+)e^{ik\cdot x}\}, 
\end{equation}
\begin{equation}
a_-(x) =\frac{1}{\sqrt{2(2{\pi})^3}}\int \frac{d^3k_+}{\sqrt{k_+}}
\frac{k_-}{k_{\bot}}{\theta}(k_+)\{
a_1(\veck_+)e^{-ik\cdot x}+
a_1^{\dagger}(\veck_+)e^{ik\cdot x}\}, 
\end{equation}
\begin{equation}
a_i(x) =\frac{1}{\sqrt{2(2{\pi})^3}}\int \frac{d^3k_+}{\sqrt{k_+}}
{\theta}(k_+){\epsilon}_i^{({2})}(k) \{
a_2(\veck_+)e^{-ik\cdot x}+ a_2^{\dagger}(\veck_+)e^{ik\cdot x}\},
\end{equation}
\begin{equation}
B(x) =\frac{1}{\sqrt{(2{\pi})^3}}\int \frac{d^3k_+}
{\sqrt{k_+}}{\theta}(k_+)k_{\bot}^2 \{ B(\veck_+)e^{-ik\cdot x}+
B^{\dagger}(\veck_+)e^{ik\cdot x}\}|_{x^-=0}, 
\end{equation}
and
\begin{equation}
C(x) =\frac{i}{\sqrt{(2{\pi})^3}}\int d^3k_+{\theta}(k_+){\sqrt{k_+}}
 \{ C(\veck_+)e^{-ik\cdot x}-
C^{\dagger}(\veck_+)e^{ik\cdot x}\}|_{x^-=0}. 
\end{equation}
We see immediately that $(3\cdot4)$ is identical with one given by
McCartor and Robertson and by Morara and Soldati and is equivalent
to the one given by Bassetto et al.  Actually, the physical
annihilation and creation operators $a_{\lambda}(\veck)$ and
$a_{\lambda}^{\dagger}(\veck)$ in the ordinary space-time coordinates are
identified with those in the light-front coordinates by
\begin{equation}
a_{\lambda}(\veck)=\sqrt{\frac{k_+}{k_0}}a_{\lambda}(\veck_+),\quad
a_{\lambda}^{\dagger}(\veck)=
\sqrt{\frac{k_+}{k_0}}a_{\lambda}^{\dagger}(\veck_+),\;({\lambda}=1,2).
\end{equation}
Consequently our $a_{\mu}-\frac{{\partial}_{\mu}}{{\partial}_-}a_-$ can be
identified with $T_{\mu}$ in the latter by changing the integration variable
from $k_+$ into $k_3$. In addition our $B$ and $C$ can be identified with
${\lambda}$ and $U$ respectively in the latter if we identify our $k_+$ with
$k_3$ in the latter.

Let us enumerate the properties of $A_{\mu}$ in $(3\cdot4)$. \\
\noindent(1) It satisfies the following 4-dimensional commutation relations
\begin{equation}
[A_{\mu}(x),\;A_{\nu}(y)]=i\{ -g_{{\mu}{\nu}}D(x-y)+
(n_{\mu}{\partial}_{\nu}+n_{\nu}{\partial}_{\mu}){\partial}_-E(x-y) \}, 
\end{equation}
where
\begin{equation}
D(x)=\frac{-i}{2(2{\pi})^3}\int\frac{d^3k_+}{k_+}{\theta}(k_+)
\{ {\rm e}^{-ik \cdot x}-
{\rm e}^{ik \cdot x} \} 
\end{equation}
and
\begin{equation}
{\partial}_-E(x)=\frac{2{\partial}_+}{{\partial}_{\bot}^2}D(x)+\frac{1}
{{\partial}_{\bot}^2}{\delta}^{(3)}(\vecx^+). 
\end{equation}
(2) It satisfies the following light-cone temporal gauge quantization
conditions:
\begin{equation}
[A_+(x),A_+(y)]|_{x^-=y^-}=[A_+(x),A_i(y)]|_{x^-=y^-}=0,\;(i=1,2) 
\end{equation}
\begin{equation}
[A_i(x),A_j(y)]|_{x^-=y^-}=-\frac{i}{2}{\delta}_{ij}({\partial}_+)^{-1}
{\delta}^{(3)}(\vecx^+-\vecy^+),\;(i,j=1,2)  
\end{equation}
\begin{equation}
[A_+(x),{\pi}^+(y)]|_{x^-=y^-}=i{\delta}^{(3)}(\vecx^+-\vecy^+), 
\end{equation}
\begin{equation}
[A_i(x),{\pi}^+(y)]|_{x^-=y^-}=\frac{i}{2}{\partial}_i({\partial}_+)^{-1}
{\delta}^{(3)}(\vecx^+-\vecy^+),\;(i=1,2) 
\end{equation}
\begin{equation}
[{\pi}^+(x),\;{\pi}^+(y)]|_{x^-=y^-}=\frac{i}{2}{\partial}_{\bot}^2
({\partial}_+)^{-1}{\delta}^{(3)}(\vecx^+-\vecy^+). 
\end{equation}
(3) The $x^-$-ordered propagator results in the ML form of propagator
\begin{eqnarray}
<{\rm{\Omega}}|\{ {\theta}(x^--y^-)
A_{\mu}(x)A_{\nu}(y)+ {\theta}(y^--x^-)A_{\nu}(y)A_{\mu}(x) \}
|{\rm{\Omega}}> \nonumber \\
=\frac{1}{(2{\pi})^4}\int d^4k\frac{i}{k^2+i{\epsilon}}\{ -g_{{\mu}{\nu}}
+\frac{n_{\mu}k_{\nu}+n_{\nu}k_{\mu}}{k_-+i{\epsilon}{\rm sgn}(k_+)} \}
{\rm e}^{-ik\cdot(x-y)}. 
\end{eqnarray}
(4) It satisfies the light-cone gauge quantization conditions in the ordinary
space time coordinates and the $x^0$-ordered propagator results in the ML form
of
propagator
\begin{eqnarray}
<{\rm{\Omega}}|\{ {\theta}(x^0-y^0)
A_{\mu}(x)A_{\nu}(y)+ {\theta}(y^0-x^0)A_{\nu}(y)A_{\mu}(x) \}
|{\rm{\Omega}}> \nonumber \\
=\frac{1}{(2{\pi})^4}\int d^4k\frac{i}{k^2+i{\epsilon}}\{ -g_{{\mu}{\nu}}
+\frac{n_{\mu}k_{\nu}+n_{\nu}k_{\mu}}{k_0-k_3+i{\epsilon}{\rm sgn}(k_3)} \}
{\rm e}^{-ik\cdot(x-y)}.
\end{eqnarray}  
\noindent(5) Translational generators can be given by integrating densities
of the canonical energy-momentum tensor over the 3-dimensional hyperplane
$x^-=$
constant.

Let us next investigate whether the properties of the light-cone axial gauge
formulation are satisfied by $(3\cdot4)$ or not. First we examine whether or
not
the light-cone axial gauge quantization conditions are satisfied
by carefully evaluating the commutation relations $(3\cdot11)$ at  $x^+=y^+.$
It
will suffice to evaluate the commutator
function ${\partial}_-E(x)$, which is rewritten as
\begin{equation}
{\partial}_-E(x)=-\frac{2}{(2{\pi})^3}\int \frac{dk_1dk_2}{k_{\bot}^2}
{\rm exp}[-i(k_1x^1+k_2x^2)]I(x^-,x^+) 
\end{equation}
where
\begin{equation}
I(x^-,x^+)=\int_0^{\infty}dk_+ \{{\rm cos}k_+x^+-{\rm cos}(k_+x^++
\frac{k_{\bot}^2}{2k_+}x^-) \}. 
\end{equation}
Note that the first and the second terms result from the ghost and physical
fields respectively and that they are ill-defined individually.
It is shown in Appendix A that $I(x^-,x^+)$ is given by
\begin{equation}
I(x^-,x^+)=\frac{\pi}{2}\{ 1+{\epsilon}(x^-){\epsilon}(x^+)\}
\sqrt{\frac{|x^-|}{2|x^+|}}\;k_{\bot}J_1(\sqrt{2|x^+x^-|}\;k_{\bot})  
\end{equation}
where $J_1(x)$ is the Bessel function of order 1. From $(3\cdot23)$ we obtain
\begin{equation}
I(x^-,x^+)|_{x^+=0}=\frac{\pi}{4}|x^-|k_{\bot}^2, \quad
{\partial}_+I(x^-,x^+)|_{x^+=0}=\frac{\pi}{2}x^-{\delta}(x^+)k_{\bot}^2
-\frac{\pi}{16}x^-|x^-|k_{\bot}^4.  
\end{equation}
It follows that
\begin{equation}
[A_+(x),A_i(y)]|_{x^+=y^+}=i{\partial}_i{\partial}_-E(x-y)|_{x^+=y^+}
=-\frac{i}{4}
|x^--y^-|{\partial}_i{\delta}^{(2)}(\vecx_{\bot}-\vecy_{\bot}), 
\end{equation}
\begin{eqnarray}
[A_+(x),A_+(y)]|_{x^+=y^+}&=&2i{\partial}_+{\partial}_-E(x-y)|_{x^+=y^+}
=-i(x^--y^-){\delta}^{(3)}(\vecx^+-\vecy^+) \nonumber \\
&-&\frac{i}{8}{\epsilon}(x^--y^-)(x^--y^-)^2{\partial}_{\bot}^2
{\delta}^{(2)}(\vecx_{\bot}-y_{\bot}).   
\end{eqnarray}
Furthermore the commutator$[A_i(x),A_j(y)]|_{x^+=y^+}$ is evaluated to be
\begin{equation}
[A_i(x),A_j(y)]|_{x^+=y^+}=i{\delta}_{ij}D(x-y)|_{x^+=y^+}=
-\frac{i}{2}{\delta}_{ij}({\partial}_-)^{-1}
{\delta}^{(3)}(\vecx^--\vecy^-).\;(i,j=1,2) 
\end{equation}
We see that these commutation relations agree with those given by Dirac's
canonical quantization procedure except the first term of $(3\cdot26)$, which
results from the fact that $I(x^-,x^+)$ has a finite discontinuity at $x^+=0$.

Next we investigate $x^+$ ordered propagator
\begin{eqnarray}
D^+_{{\mu}{\nu}}(x-y)&=&<{\rm{\Omega}}|\{ {\theta}(x^+-y^+)
A_{\mu}(x)A_{\nu}(y)+ {\theta}(y^+-x^+)A_{\nu}(y)A_{\mu}(x) \}
|{\rm{\Omega}}> \nonumber \\
&=&\frac{1}{(2{\pi})^4}\int d^4q\;D^+_{{\mu}{\nu}}(q){\rm e}^{-iq\cdot (x-y)}.
\end{eqnarray}  
Note that $B$ and $C$ are zero-norm fields so they have nonvanishing
contributions for three cases, namely for ${\mu}=+,{\nu}=i ;
{\mu}=i,{\nu}=+$ and ${\mu}={\nu}=+$. In case that ${\mu}=i$ and ${\nu}=j$
there arises no
problem and we have
\begin{equation}
D^+_{ij}(q)=\frac{i{\delta}_{ij}}{ q^2+i{\epsilon}}.
\end{equation}  
When ${\mu}=+$ and ${\nu}=i$ or ${\mu}=i$ and ${\nu}=+$, we obtain
\begin{eqnarray}
D^+_{+i}(q)=D^+_{i+}(q)&=&\frac{q_i}{q_-}\cdot \frac{i}{q^2+i{\epsilon}}
-\frac{iq_i}{ q_{\bot}^2}(-i{\pi}){\rm sgn}(q_+){\delta}(q_-) \nonumber \\
&=&\frac{i}{q^2+i{\epsilon}}\cdot \frac{q_i}{q_-+i{\epsilon}{\rm sgn}(q_+)}.
\end{eqnarray}  
We see that the second term, which is the contribution from the ghost fields,
contributes the ${\delta}$ function part of the ML prescribed propagator.

In case that ${\mu}={\nu}=+$, we obtain
\begin{equation}
D^+_{++}(q)=\frac{q_{\bot}^2}{q_-^2}\cdot \frac{i}{q^2+i{\epsilon}}
-\frac{2{\delta}(q_-)}{ q_{\bot}^2}\int_0^{\infty}dk_+
(\frac{ik_+}{q_+-k_++i{\epsilon}}-\frac{ik_+}{q_++k_+-i{\epsilon}}) .
\end{equation}  
As was noticed by Morara and Soldati,$^{11)}$ the second term diverges
linearly. This implies that if the ghost fields really regularize
$D^+_{++}(x-y)$, a linear divergence has to appear from the physical
contribution so as to be canceled.  As a matter of fact we see
that if we rewrite the physical contribution as
\begin{equation}
\frac{q_{\bot}^2}{q_-^2}\cdot \frac{i}{q^2+i{\epsilon}}
=\frac{2q_+}{q_-}\cdot \frac{i}{q^2+i{\epsilon}}-\frac{i}{q_-^2},
\end{equation}  
then the second term $\frac{i}{q_-^2}$ gives rise to a linear
divergence as follows
\begin{equation}
\int_{-\infty}^{\infty}dq_-\frac{1}{q_-^2}{\rm e}^{iq_-x^-}
=2\int_0^{\infty}dq_-\frac{{\rm cos}q_-x^-}{q_-^2}
=2\int_0^{\infty}dq_-\frac{1}{q_-^2}-{\pi}|x^-|
\end{equation}
when $D^+_{++}(x)$ is restored by inverse Fourier transform.  Furthermore by
changing the integration variable from $k_+$ into
$q_-=\frac{q_{\bot}^2}{2k_+}$, we see that it is canceled by the linear
divergence due to the ghost fields as follows
\begin{eqnarray}
&\int_{-{\infty}}^{\infty}dq_-\frac{4i}{q_{\bot}^2}{\delta}(q_-)
(\int_0^{\infty}dk_+){\rm e}^{-iq_-x^-}-2i\int_0^{\infty}\frac{dq_-}{q_-^2}&
\nonumber \\
&=2i(\int_0^{\infty}\frac{2dk_+}{q_{\bot}^2}-\int_0^{\infty}\frac{dq_-}{q_-^2})
=2i(\int_0^{\infty}\frac{dq_-}{q_-^2}-\int_0^{\infty}\frac{dq_-}{q_-^2})
=0.& 
\end{eqnarray}
To the best of our knowledge this point has been overlooked so far.
We verify in Appendix B that  linear divergences do
not appear from $D^+_{++}(x)$ if we carry out $k_+$-integrations of the
physical and ghost contributions simultaneously and in advance of Fourier
transformation with respect $x^+$. This verifies that the linear divergences
arising from the ghost and physical contributions cancel each other.
Therefore, although changing the order of integrations prevents us from
carrying out the Fourier transformations to obtain a closed form expression for
$D^+_{++}(q)$, we are justified in neglecting the linear divergence arising
from the ghost contribution as follows
\begin{eqnarray}
&D^+_{++}(q)=\frac{2q_+}{q_-}\cdot \frac{i}{q^2+i{\epsilon}}
-\frac{2iq_+}{ q_{\bot}^2}(-i{\pi}){\varepsilon}(q_+){\delta}(q_-)
-\frac{i}{2}\{ \frac{1}{(q_-+i{\epsilon})^2}+\frac{1}{(q_--i{\epsilon})^2} \}&
\nonumber \\
&=\frac{i}{q^2+i{\epsilon}}\cdot
\frac{2q_+}{q_-+i{\epsilon}{\varepsilon}(q_+)}-\frac{i}{2}\{
\frac{1}{(q_-+i{\epsilon})^2}+\frac{1}{(q_--i{\epsilon})^2} \}.&
\end{eqnarray}  
 Here the finite contact term is obtained as a result of Fourier
transformation of $|x^-|$. The appearance of contact terms implies the
existence of Coulomb-like counter terms in the interaction Hamiltonian.
We leave the specification of such terms  for subsequent studies.

\section{Concluding Remarks}
In this paper we have analyzed further the problem of the canonical derivation
of the propagator in gauge theories, both in the light-cone representation and
in the equal-time representation. We have emphasized the indispensability of
ghost fields in the light-cone gauge formulation of gauge fields.
In particular, we have found that the ghost fields are indispensable if we are
to
have well-defined antiderivatives and to properly regularize the most
singular component of the naive gauge field propagator. We have also shown
 that
the ML form of propagator is obtained in a consistent temporal gauge
formulation
of gauge fields in the ordinary space-time coordinates if a suitable vacuum is
chosen.

What we have not completely understood is the discrepancy observed in the
commutator $[A_+(x),A_+(y)]|_{x^+=y^+}$, in which we find
an extra ${\delta}(x^+-y^+)$-type singularity. It seems that this discrepancy
indicates that the light-cone limits are not the same and that the light-cone
axial gauge formulation will require the elimination of such  singularities.
We have also left completing the Fourier transform of the (formally)
most singular component of the gauge field propagator for subsequent studies.

Finally we point out that our approach may provide an easier way to construct
perturbation theories of interacting gauge fields in the light-cone temporal
gauge than that of Morara and Soldati$^{11)}$ , because  temporal
gauge Lagrangians are regular in the auxiliary coordinates. We also leave this
task for subsequent studies.

\section*{Acknowledgements}

The work of one of us (GMc) was supported by grants from the U.S. Department of
Energy.

\appendix
\section{ Verification of $(3\cdot24)$}
With the help of the addition theorem of the trigonometric cosine function,
we can decompose $I(x^-,x^+)$ as a sum of two well-defined integrals
\begin{equation}
I(x^-,x^+)=I^{(1)}(x^-,x^+)+I^{(2)}(x^-,x^+)
\end{equation}
where
\begin{equation}
I^{(1)}(x^-,x^+)=\int_0^{\infty}dk_+{\rm sin}k_+x^+ \cdot{\rm sin}
\frac{k_{\bot}^2x^-}{2k_+},
\end{equation}
\begin{equation}
I^{(2)}(x^-,x^+)=\int_0^{\infty}dk_+{\rm cos}k_+x^+ \cdot(1-{\rm cos}
\frac{k_{\bot}^2x^-}{2k_+}).
\end{equation}
We see that $I^{(1)}(x^-,x^+)$ is an odd function of $x^+$ and of $x^-$
and becomes trivially zero at $x^+=0$ and/or at $x^-=0$. We also see that
in the case that $x^+>0$ and $x^->0$, an explicit expression of
$I^{(1)}(x^-,x^+)$
is known from an integral formula$^{18)}$
\begin{equation}
\int_0^{\infty}dk\;{\rm sin}ak \cdot{\rm sin}\frac{b}{k}=
\frac{\pi}{2}\sqrt{\frac{b}{a}}J_1(2\sqrt{ab}),\;(a>0,b>0)
\end{equation}
where $J_1$ stands for the Bessel function of order 1. Therefore
we immediately obtain
\begin{equation}
I^{(1)}(x^-,x^+)=\frac{\pi}{2}{\epsilon}(x^-){\epsilon}(x^+)
k_{\bot}\sqrt{\frac{|x^-|}{2|x^+|}}J_1(k_{\bot}\sqrt{2|x^-x^+|}).
\end{equation}
Because of the sign factor ${\epsilon}(x^+), I^{(1)}(x^-,x^+)$ gives a finite
discontinuity at $x^+=0$.

Next we show that an explicit expression for $I^{(2)}(x^-,x^+),$ which is an
even function of $x^-$ and of $x^+$, results from the same formula
$(A\cdot4)$. In case that $x^+{\neq}0$, integrating by parts and then
changing integration variable from $k_+$ to $k_-=\frac{k_{\bot}^2}{2k_+}$
results in
\begin{eqnarray}
I^{(2)}(x^-,x^+)&=&\frac{x^-}{2x^+}k_{\bot}^2\int_0^{\infty}\frac{dk_+}
{k_+^2}{\rm sin}k_+x^+ \cdot{\rm sin}\frac{k_{\bot}^2x^-}{2k_+}
=\frac{x^-}{x^+}\int_0^{\infty}dk_-{\rm sin}k_-x^- \cdot
{\rm sin}\frac{k_{\bot}^2x^+}{2k_-} \nonumber \\
&=&k_{\bot}\frac{\pi}{2}\sqrt{\frac{|x^-|}{2|x^+|}}
J_1(k_{\bot}\sqrt{2|x^-x^+|}). 
\end{eqnarray}
The value at $x^+=0$ is calculated as follows
\begin{eqnarray}
I^{(2)}(x^-,x^+)|_{x^+=0}&=&\int_0^{\infty}dk_+(1-
{\rm cos}\frac{k_{\bot}^2x^-}{2k_+})
=\frac{k_{\bot}^2}{2}\int_0^{\infty}dk_-\frac{1-{\rm cos}k_-x^-}{k_-^2}
 \nonumber \\
&=&\frac{k_{\bot}^2x^-}{2}\int_0^{\infty}dk_-\frac{{\rm sin}k_-x^-}{k_-}
=\frac{\pi}{4}k_{\bot}^2|x^-|, 
\end{eqnarray}
which turns out to be the limit of $(A\cdot6)$ as $x^+ \to0$. Consequently
$I^{(2)}(x^-,x^+)$ is known to be a continuous function of $x^-$ and $x^+$.
Substituting $(A\cdot5)$ and $(A\cdot6)$ into $(A\cdot1)$ yields $(3\cdot24)$.

\section{Disappearance of the linear divergence from $D^+_{++}(x)$}
To demonstrate that $D^+_{++}(q)$ possesses no linear divergences owing to
the ghost fields, we carry out the $k_+$-integration in
\begin{equation}
D^+_{++}(x)=\frac{2}{(2{\pi})^3}\int d^3k_+ \frac{k_+}{k_{\bot}^2}{\theta}(k_+)
\{ {\theta}(x^+)({\rm e}^{-ik\cdot x}-{\rm e}^{-ik\cdot x}|_{x^-=0})
+{\theta}(-x^+)({\rm e}^{ik\cdot x}-{\rm e}^{ik\cdot x}|_{x^-=0}) \}
\end{equation}
in advance of Fourier integrations yielding $D^+_{++}(q).$ We can then rewrite
$D^+_{++}(x)$ in the following form
\begin{equation}
D^+_{++}(x)=\frac{2}{(2{\pi})^3}\int\frac{dk_1dk_2}{k_{\bot}^2}
{\rm exp}[-i(k_1x^1+k_2x^2)]\{ {\partial}_+J(x^-,x^+)-i{\epsilon}(x^+)
{\partial}_+I(x^-,x^+) \}
\end{equation}
where
\begin{equation}
J(x^-,x^+)=\int_0^{\infty}dk_+ \{ {\rm sin}(k_+x^++\frac{k_{\bot}^2x^-}{2k_+})
-{\rm sin}k_+x^+ \}
\end{equation}
and $I(x^-,x^+)$ is the function given in Appendix A. We see that
owing to the second term, which comes from the ghost contribution, the
integrand of
$J(x^-,x^+)$ behaves like $\frac{{\rm cos}k_+x^+}{k_+}$ as $k_+\to{\infty}$,
which verifies that $J(x^-,x^+)$ is well-defined when $x^+{\neq}0$.

It is useful to illustrate how the ghost contribution gives rise to a
linear divergence. Applying the distribution procedure we obtain
\begin{equation}
\int_0^{\infty}{\rm sin}k_+x^+=\frac{-i}{2}\int_{-\infty}^{\infty}
{\epsilon}(k_+){\rm e}^{ik_+x^+}={\rm P}\frac{1}{x^+}
\end{equation}
and hence
\begin{eqnarray}
\int_{-\infty}^{\infty}dx^+{\rm e}^{iq_+x^+}{\partial}_+
(\int_0^{\infty}{\rm sin}k_+x^+)&=&-\int_{-\infty}^{\infty}dx^+
\frac{1}{(x^+)^2}{\rm e}^{iq_+x^+}=-2\int_0^{\infty}dx^+\frac{{\rm cos}q_+x^+}
{(x^+)^2} \nonumber \\
&=&-2\int_0^{\infty}dx^+\frac{1}{(x^+)^2}+{\pi}|q_+|. 
\end{eqnarray}
We see that the last term diverges linearly. Furthermore we notice that
showing that $J(x^-,x^+)$ does not possess any terms tending to $\frac{1}{x^+}$
as $x^+{\to}0$ verifies the disappearance of linear divergences from
$D^+_{++}(x)$.
Note that $I(x^-,x^+)$ has no such terms, as is seen from $(3\cdot25)$.

We can find an explicit expression for $J(x^-,x^+)$ by making use of the
${\mu}{\to}1-{\tiny 0}$ limits of the following integral formulas$^{19)}$
\begin{equation}
\int_0^{\infty}dkk^{{\mu}-1}{\rm sin}(ak-\frac{b}{k})=
2\left(\frac{b}{a}\right)^{\frac{\mu}{2}}{\rm sin}\frac{{\mu}{\pi}}{2}K_{\mu}
(2\sqrt{ab}),
\end{equation}
\begin{equation}
\int_0^{\infty}dkk^{{\mu}-1}{\rm sin}(ak+\frac{b}{k})={\pi}
\left(\frac{b}{a}\right)^{\frac{\mu}{2}}
\{ {\rm cos}\frac{{\mu}{\pi}}{2}J_{\mu}(2\sqrt{ab})-
{\rm sin}\frac{{\mu}{\pi}}{2}N_{\mu}(2\sqrt{ab}) \}
\end{equation}
where $a\!>\!0,b\!>\!0,|{\rm Re}{\mu}|\!<\!1$ and $K_{\mu},J_{\mu}$ and
$N_{\mu}$ are Bessel functions. We are justified in using the limits to
calculate $J(x^-,x^+)$, because $J(x^-,x^+)$ is decomposed as a difference
of the two integrations and because the difference is well-defined so that
it is independent of the regularizations. In the limit we have
\begin{equation}
\int_0^{\infty}dk\;{\rm sin}(ak-\frac{b}{k})=
2\sqrt{\frac{b}{a}}K_1(2\sqrt{ab}),
\end{equation}
\begin{equation}
\int_0^{\infty}dk\;{\rm sin}(ak+\frac{b}{k})=-{\pi}
\sqrt{\frac{b}{a}}N_1(2\sqrt{ab}).
\end{equation}
Furthermore, taking the $b{\to}0{\tiny +}$ limit of $(B\cdot8)$ and
$(B\cdot9)$ yields
\begin{equation}
\int_0^{\infty}dk\;{\rm sin}ak=\frac{1}{a}
\end{equation}
which agrees with $(B\cdot4)$. It follows from $(B\cdot8){\sim}(B\cdot10)$ that
\begin{eqnarray}
&{}&J(x^-,x^+)=\frac{{\epsilon}(x^+)+{\epsilon}(x^-)}{2}k_{\bot}
\sqrt{\frac{|x^-|}{2|x^+|}}\{ -{\pi}N_1(k_{\bot}\sqrt{2|x^-x^+|})-
\frac{1}{k_{\bot}}\sqrt{\frac{2}{|x^-x^+|}} \} \nonumber \\
&+&\frac{{\epsilon}(x^+)-{\epsilon}(x^-)}{2}k_{\bot}
\sqrt{\frac{|x^-|}{2|x^+|}}\{ 2K_1(k_{\bot}\sqrt{2|x^-x^+|})-
\frac{1}{k_{\bot}}\sqrt{\frac{2}{|x^-x^+|}} \}.
\end{eqnarray}
This shows that the leading term for small $|x^+|$ is $-\frac{1}{2}x^-
{\rm log}|x^+|$ so that no linear divergences arise in the
Fourier transform of ${\partial}_+J(x^-,x^+)$.

\end{document}